\documentclass[onecolumn]{mn2e}
\usepackage{times}

\input{psfig.sty}

\newif\ifAMStwofonts

\title[Iron line profiles and inner boundary condition in accretion flows]
        {Iron K$\alpha$ line profiles and the
         inner boundary condition of accretion flows
}
\author[A. Merloni \& A. C. Fabian]
        { A. Merloni$^1$ and  A. C. Fabian$^2$
\\$^1$Max-Planck-Institut f\"ur Astrophysik,
        Karl-Schwarzschild-Strasse 1, D-85741, Garching, Germany 
\\$^2$Institute of Astronomy, Madingley Road, CB3 0HA, Cambridge
}

\date{}

\begin{document}

\maketitle

\label{firstpage}

\begin{abstract}
Recent X-ray observations have shown evidence for exceptionally broad
and skewed iron K$\alpha$ emission lines from several accreting black
hole systems. The lines are assumed to be due to fluorescence of the
accretion disk illuminated 
by a surrounding corona and require a steep emissivity
profile increasing in to the innermost radius. This appears to
question both standard accretion disc theory and the zero torque
assumption for the inner boundary condition, both of which predict a
much less extreme profile. Instead it argues that a torque may be
present due to magnetic coupling with matter in the plunging region or
even to the spinning black hole itself. Discussion so far has centered
on the torque acting on the disc. However the crucial determinant of
the iron line profile is the radial variation of the power radiated in
the corona. Here we study the effects of different inner boundary
conditions on the coronal emissivity and on the profiles of the
observable Fe K$\alpha$ lines. 
 We argue that in the extreme case where a
prominent high redshift component of the iron  line is
detected, requiring a steep emissivity profile in the innermost part
and a flatter one outside, energy from the gas
plunging into the black hole is being fed directly to the corona.
\end{abstract}


\section{Introduction}
The much improved sensitivity of the X-ray satellite {\it XMM Newton}
is quickly expanding our capability of performing accurate X-ray
spectroscopy of accretion powered systems. In several recent
observations of accreting black holes, extremely broadened and
redshifted iron K$\alpha$ lines have been detected. Wilms et al.
(2001) presented the observations of the nearby Seyfert 1 galaxy
MCG--6-30-15 in a low-luminosity state and deduced from the iron
K$\alpha$ line profile that the reprocessing material must be located
in the very innermost part of the accretion flow. Similarly, Miller et
al. (2002), found a very broad, skewed Fe K$\alpha$ emission line in
the galactic black hole XTE J1650-500 in its {\it very high} state. In
both cases, not only is the large observed redshift considered as a
signature of a rapidly rotating black hole (so that the innermost
stable orbit is close to the event horizon and the accretion disc can
penetrate much deeper in the potential well), but also the disc
emissivity needs to be strongly peaked towards the inner boundary in
order to fit the observed line profile. For simplicity, the disc
emissivity is usually assumed to have a power-law profile $F(r)
\propto r^{-\beta}$ (where $r=R/R_{\rm S}=Rc^2/2GM$ is the distance
from the center in units of Schwarzschild radii), and the index of such power-law is generally
assumed to be given by the accretion disc emissivity index in a
standard geometrically thin and optically thick disc \cite{ss73}. In
this case we expect $\beta<3$. The fact that both Wilms et al. (2001)
and Miller et al. (2002) require a larger value of $\beta$ to fit the
line profile has prompted speculation that, in the inner disc
region, additional energy dissipation must be taking place
\cite{ak00,li02}. This might be due to magnetic torque exerted by the
accreting material in the plunging region at the inner disc boundary
\cite{gam99,kro99} or to magnetic extraction of the spin energy of the
black hole \cite{li02} through magnetic field lines threading the
ergosphere and the accretion disc.
Such findings have been confirmed by the analysis of a long 
observation of MCG--6-30-15 with {\it XMM-Newton} reported by Fabian
et al. (2002). These authors found that ``the strong, skewed iron line
is clearly detected and is well characterized by a steep emissivity
profile within $6GM/c^2$ and a flatter profile beyond.''

Here we first make the point that, 
 in the framework of standard, relativistic accretion
disc models with zero torque at the inner boundary, the emissivity
profile of the hard illuminating X-rays is more centrally peaked than
that of the usually assumed standard accretion disc.
This is a general
property of the energy deposition rate in a magnetic corona above
standard accretion discs. Furthermore, the quality of the X-ray
spectroscopic data is such that, using our coronal models, we can
start considering the nature of the inner boundary condition for
relativistic accretion flows. Using a simple analytic formulation,
first introduced by Agol \& Krolik (2000) to include magnetic stresses
at the marginally stable orbit in standard disc equations, we can
estimate the effect of a modified inner boundary condition on the
observed iron line profiles. In particular, we consider the three
following possibilities:
\begin{itemize}
\item[1)]{There is no net torque at the inner boundary (no-torque case;
    NT);}
\item[2)]{Magnetic field lines couple the plunging region to, 
and exert stress on, the  {\it cold, geometrically thin disc} only
(disc-torque; DT);}
\item[3)]{Magnetic field lines couple the plunging region to the
    {\it vertically extended, magnetically dominated corona} only
    (coronal-thread; CT).}
\end{itemize}
We show that, while in the NT case the
coronal emission is more centrally peaked than the disc one (due to
the radial dependence of the fraction of energy dissipated there), 
the effect is greatly enhanced in the CT case, with the
strongest deviation from the standard emissivity profile appearing in
the innermost coronal region. 
Although the emissivity profiles again become more centrally peaked
when the torque is applied directly to the thin disc (DT), the coronal
power is much reduced, and we expect the emission to be completely
dominated by the quasi-thermal radiation coming from the optically
thick disc.

The structure of the paper is the following: 
in section 2 we solve the fully general relativistic structure
equations for standard geometrically thin and optically thick
accretion discs, coupled to a magnetically dominated corona (with the
coupling discussed in Merloni, 2003, hereafter M03) and opportunely
modified by the inclusion of a term describing the additional
dissipation from magnetic stresses at the inner boundary either in the
disc or in the corona. In such a way we are able to compute
the radial profiles of the hard coronal radiation that illuminates the
cold disc and is responsible for the fluorescent iron K$\alpha$ line
production (section~\ref{sec_torque}). Then, by means of a simple
 broken power-law approximation for the coronal emissivity, 
in section~\ref{sec_lines},
we simulate and compare the expected line profiles for the different
cases. In section~\ref{sec_discuss} we discuss a number of possible
shortcomings of the model we have adopted, and in
section~\ref{sec_conc} we draw our conclusions.

\begin{figure*}
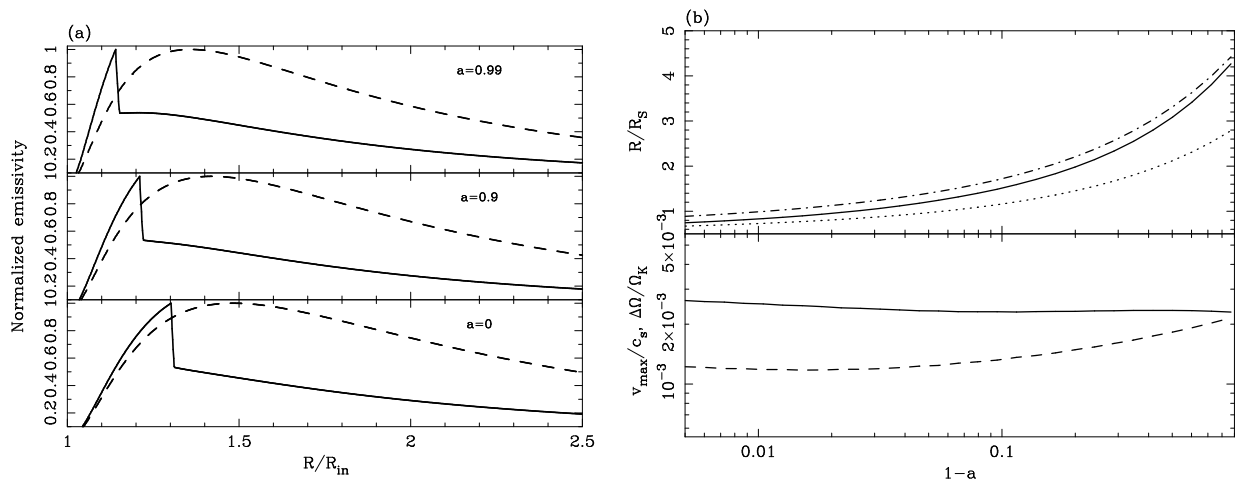

\begin{tabular}{cc} 
\psfig{angle=270,width=0.45\textwidth,file=r_f_nt.ps}
&\psfig{angle=270,width=0.45\textwidth,file=r_v_nt.ps}\\ 
\end{tabular}
\caption{On the left (panel a), coronal emissivity profiles 
$F_{\rm c}$ (solid lines) 
and corresponding standard disc 
emissivities $Q$ (dashed lines), both normalized
to their maximum level, as a function of the disc
radius, in units of the inner accretion disc radius (fixed at the
  marginally stable orbit) for three different values of the
dimensionless black hole angular momentum: $a=0$;  
$a=0.9$  and
$a=0.99$, from bottom to top. On the right (panel b), the upper panel shows the location of the innermost stable
  orbit (dotted line), the radius of maximal coronal emissivity
  (solid line) and that of maximal disc emissivity (dashed line), all
  in units of Schwarzschild radii, as 
  functions of the dimensionless black hole spin parameter $a$, for the
  case of no torque at the inner boundary. In the lower panel the
  corresponding radial velocity of the gas (in units of the local
  sound speed) and the relative departure of the gas angular velocity
  from the Keplerian value (assumed in the computations), 
at the radius of maximal
  coronal emissivity are shown.}
\label{gas_mdot01}
\end{figure*}

\section{Centrally peaked coronal emissivity: the no-torque case}
\label{centr}

In Merloni (2003; see also Merloni \& Fabian 2002), we have shown how it is possible to build
a physically self-consistent analytical model for a thin accretion disc--corona
system by
solving algebraic equations for the fraction of power released into
the corona as a function of the distance from the central
(non-rotating) black hole.
Here we extend such a treatment to the general relativistic case of
accretion discs around rotating black holes by calculating the
disc--corona structure equations (presented
in the Appendix) which are a generalization of the classical ones first
derived by Novikov \& Thorne (1973), and take into account both the
effects of the presence of a corona on the disc structure and those of the
different viscosity prescription on the coronal emissivity profiles.

The nature of viscosity in accretion flows is still one of the most
fundamental open questions of black hole astrophysics. At present,
magneto-rotational instability \cite{bh91,bh98} is favoured as the
primary source of the turbulent viscosity needed to explain the
luminosities of accreting black holes. For this reason we expect that the
turbulent magnetic stresses responsible for the angular momentum
transport in the disc are given by
\begin{equation}
\rho W_{R\phi} \simeq \frac{\langle B_R \times B_{\phi}\rangle}{8 \pi}
\simeq P_{\rm mag}.
\end{equation}
Once the relationship between magnetic and disc pressure
(given either by the gas or by the radiation) is established, the
accretion disc structure can be fully described.
This is a long standing issue in accretion disc theory and, in the
framework of MRI-driven turbulent viscosity models, can be translated
into the uncertainty about the mechanisms by, and the level at, which the
disc magnetic field saturates. Numerical simulations
are crucial to test and verify different hypothesis on the outcome of the
instability and the final scaling of the magnetic pressure. 
However, full three dimensional MHD 
simulations of accretion flows that include radiative transfer and is
capable to treat radiation-pressure-dominated solutions (on the line
of those presented in Turner, Stone \& Sano, 2002) are needed, in
order to answer the question of the viscosity scaling for luminous
thin discs. 
On the other hand, by means of a more treatable analytic approach,
 Merloni (2003) has shown that, as the 
growth rate of the MRI in radiation-pressure-dominated discs is
reduced by a factor $c_{\rm g}/v_{\rm A}$ \cite{bs01,tss02}, 
where $c_{\rm g}$ is the
gas sound speed and $v_{\rm A}$ is the Alfv\`en speed, and if the 
 main saturation mechanism for the magnetic field is vertical buoyancy
(as would be required in order to produce powerful coronae), then 
(see also Taam \& Lin, 1984; Burm, 1985)
\begin{equation}
\label{szu}
P_{\rm mag}=\alpha_0 \sqrt{P_{\rm gas} P_{\rm tot}},
\end{equation}
where $\alpha_0$ is a constant (not necessarily smaller than unity,
see M03).
Then it is easy to show that,
locally in the disc, the fraction of power released into the corona 
\cite{hm91,sz94} is given by
\begin{equation}
\label{eq_fr}
f(R) \simeq \sqrt{2\alpha_0} \left(1+\frac{P_{\rm
      rad}(R)}{P_{\rm gas}(R)}\right)^{-1/4}.
\end{equation}
The first important implication of the above result is that 
the coronal fraction $f$ is reduced when
radiation pressure dominates\footnote{Indeed, many of the results that
  are discussed here are valid in a more general framework than that
  proposed by M03. It is in fact sufficient that the
  magnetic pressure, Eq. (\ref{szu}), be not proportional to the
  total pressure only in order for a dependency of $f$ on the
  pressures ratio to be introduced.}. As radiation pressure is more and
more dominant with increasing accretion rates, a
dependency of the spectral properties of an accreting source (that are
very sensitive to the fraction of power released in the hot,
optically thin corona) on the accretion rate is introduced. 
This may be relevant, for example, for
understanding spectral transition in galactic black hole binaries
\cite{mf02}.

\begin{figure*}
\psfig{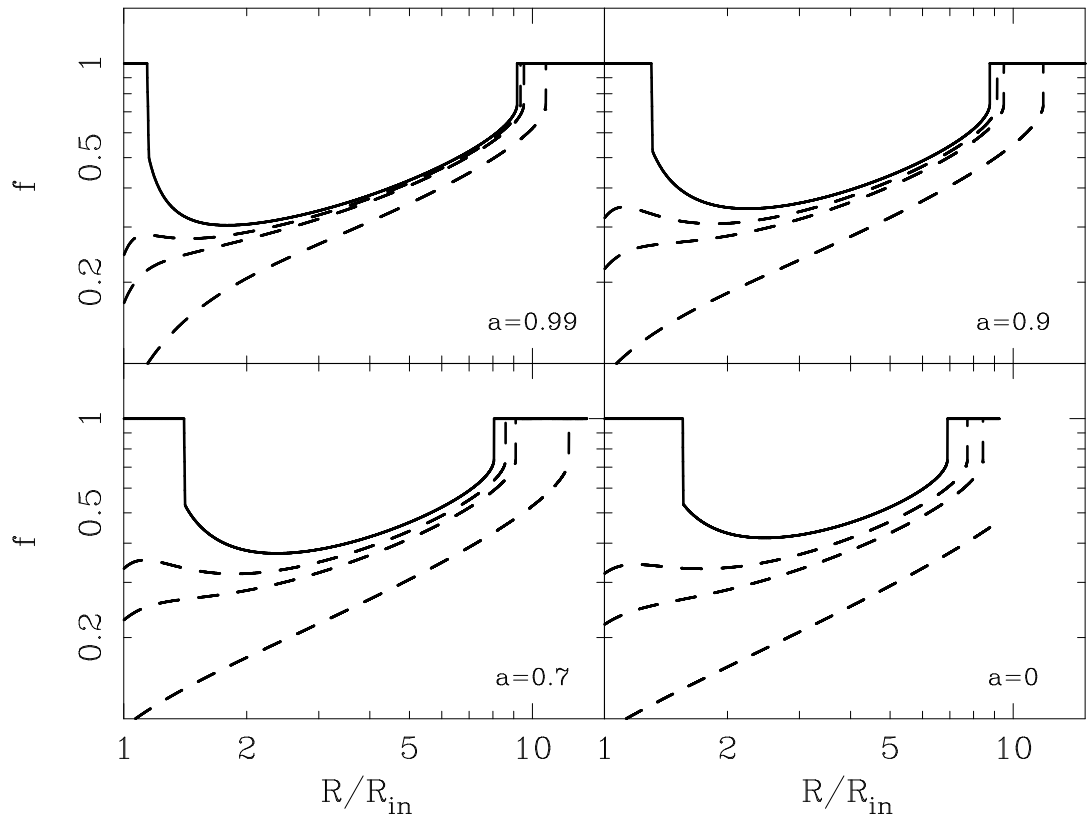}
\caption{The profiles of the coronal fraction $f(R)$ for $\Delta\epsilon=0$ (NT, thick solid lines) and 
for the case of disc torque (DT) with $\Delta\epsilon=0.05, 0.1, 0.5$ (dashed lines, $\Delta \epsilon$
increasing from top to bottom).
The dimensionless black hole spin $a=0.99, 0.9, 0.7 ,0$ is indicated
in the different panels. 
The dimensionless accretion rate, defined as the ratio of the emerging
(accretion) bolometric luminosity to the Eddington one has been kept fixed to 
$\dot m \equiv L_{\rm b,acc}/L_{\rm Edd} = \epsilon(a) \dot M_0 c^2/ L_{\rm Edd}=
0.03$, where $\dot M_0$ is the constant external accretion rate and 
$\epsilon(a)$ is the standard radiative efficiency computed in terms of
binding energy at $R_{\rm in}=R_{\rm ms}$, which is a function of the
dimensionless black hole spin $a$.}
\label{prof_dt}
\end{figure*}

At any given accretion rate
the coronal emissivity profile $F_{\rm c}(R)$, which determines the
intensity of the hard X-ray radiation illuminating the disc at every
radius, is given by
\begin{equation}
F_{\rm c}(R)=f(R)Q(R)=\frac{3 G M \dot M_0f(R)}{8 \pi R^3}\frac{{\cal Q}
  {\cal B}}{ {\cal C}^{1/2}}\; {\rm ergs}\; {\rm cm}^{-2} \; {\rm s}^{-1},
\end{equation} 
where ${\cal Q}$, ${\cal B}$ and  ${\cal C}$ are the relativistic
correction factors of Novikov and Thorne (1973), approaching unity at
large distances (see Appendix). One of the key feature of such a 
relativistic solution is the `no-torque at the inner boundary'  
(in the following, NT) condition. 
Such a condition ensures that the disc radiative flux
vanishes at the inner boundary, and this in turn implies that 
the ratio of radiation to gas pressure decreases as
the inner boundary is approached from outside. In fact, the innermost
part of a standard geometrically thin and optically thick disc is
always {\it gas pressure dominated} if the NT condition is satisfied, and the
coronal emissivity should therefore be strongly peaked towards the center.

In order to show this, we have calculated the coronal emissivity 
profiles for a supermassive
black hole ($m=10^8$) accreting at about three per cent of the Eddington rate 
(a rate such that at least a portion of the inner disc is dominated by
radiation pressure) and with 
$\alpha_0=0.49$, for different values of the black hole spin.
We solve the relativistic disc structure equations, modified by the
presence of the corona (see Eq. A2-A11), with the disc-corona coupling
given by Eq.~(\ref{eq_fr}).
An example of the results  
is shown in Fig.~\ref{gas_mdot01}a. 
The
emissivity profile of the corona $F_{\rm c}$ is much more strongly 
peaked than the corresponding disc emissivity profile
$Q$. This effect does not depends on neglecting the effects of
pressure gradients near the inner boundary.
Indeed, as it is also clear from Fig.~\ref{gas_mdot01}a, the peak of the coronal
emissivity is {\it not}  located near the innermost stable orbit, if
``near'' here means so close that substantial deviation from the values of the
physical quantities computed with the Novikov-Thorne equations 
should be considered.
In fact, the coronal emissivity profile peaks 
close to the radius where the gas pressure is equal to the radiation
pressure. 
Calculations done for high accretion rate discs, including the effects
of pressure gradients, show  that the ratio of
gas to total pressure increases substantially when going towards the
center \cite{abr88}, from $r/r_{\rm in}=5/3$ to
$r/r_{\rm in}=1$ (see e.g. Fig. 6 of  Muchotrzeb \& Paczynski, 1982,
or Fig. 7 of Abramowicz et al. 1988), 
even for much higher accretion rates  than those we are considering.
Indeed, we can compute from our numerical solutions the ratio of the
radial velocity at the location of the maximal coronal emissivity,
$v_r(r_{\rm max})\equiv v_{\rm max}$, and compare it with the local
sound speed, to have a measure of the distance between $r_{\rm max}$
and the sonic point. Moreover, the error we make by neglecting
 pressure gradients
can be estimated calculating the relative departure of the gas angular velocity
from the Keplerian value at $r=r_{\rm max}$, from 
the momentum equation in the $r$ direction:
\begin{equation}
\frac{\Delta\Omega}{\Omega_{\rm K}}\equiv
\frac{(\Omega^2-\Omega_{\rm
  K}^2)^{1/2}}{\Omega_{\rm K}}=
\frac{1}{r \Omega_{\rm K}} \left(\frac{1}{\rho}\frac{dP}{dr}+v_r\frac{dv_r}{dr}\right)
\end{equation}

In Fig.~\ref{gas_mdot01}b we show the location of the inner disc
radius, of the radius of maximal coronal emissivity and of maximal
disc emissivity without corona (upper panel) together with
the values of $v_{\rm max}/c_{\rm s}$ and $\Delta \Omega / \Omega_{\rm
  K}$ (lower panel), as a function of the dimensionless black hole
angular momentum $a$ for the NT case. 
At the location of the maximal
coronal emissivity, the radial flow is about three orders of 
magnitude subsonic and the error we make in neglecting pressure
gradient and advective terms is a fraction of a percent.

The coronal emissivity profiles shown in
Figure~\ref{gas_mdot01}a represent
an upper limit on the central concentration of hard illuminating
X-rays that can be achieved without the need of any kind of energy
extraction from matter inside the marginally stable orbit. 
Comparison of the calculated emissivity
profiles with those needed to explain the broad relativistic lines
observed in the spectra of the Seyfert 1 galaxy MCG--6-30-15 
\cite{wil01,fab02}, reveals
that the radial extent of the region with emissivity index $\beta>3$
(the ``core'' region, see below) is probably too small in the NT
scenario to account for the observations.
In the next section we investigate the effects of applying a torque at
the inner disc boundary on the coronal emissivity profiles.

\section{Torque at the inner disc boundary: thin vs. thick discs}
\label{sec_torque}
The original idea that magnetic field inside the radius of marginal
stability ($R_{\rm ms}$) may alter the dynamics of accreting material and
act to effectively
torque the disc at its inner boundary 
was already mentioned in the early works on black hole
accretion discs (e.g. in Page \& Thorne, 1974), although
the standard hypothesis, in the following three decades of accretion
disc research, has been that of no-torque at the inner boundary
(see Abramowicz \& Kato, 1989, and references therein). When the NT
inner boundary condition is adopted, the maximal energy per unit mass 
available to be radiated by the accretion disc is uniquely determined
by the binding energy at the innermost stable orbit and this fixes the
efficiency of accretion. On the other hand, if dynamically important
magnetic fields in the plunging region exert significant stresses on
the disc (disc torque condition, in the following DT), 
the available energy per unit mass increases and the
efficiency may even reach values greater than unity, if magnetic stress 
can be exerted on the disc from material that has plunged into the
ergosphere, thus effectively tapping the rotational energy of the
black hole \cite{kro99,gam99,ak00}. As a consequence, 
the nature of the inner disc boundary
condition, and in particular the extent of the torque exerted by the
plunging gas on the accretion flow, 
depends on the complicated nature of the (turbulent) angular
momentum transport in the disc and in particular on the level of the
saturation magnetic field inside the disc \cite{ra01,kh02}.
The recent MHD simulations of Hawley and Krolik (2002), which
are the ones with the best resolution to date, seem to confirm the idea that
 magnetized discs do not satisfy the no-torque at the inner
boundary condition,
showing that the stress generated by the MHD turbulence 
does not vanish near the marginally stable orbit.

There is in fact a contrasting argument in such a debate. 
Indeed, all the simulations that have shown signs of magnetic stress
inside the marginally stable orbit were non-radiative, and involved
moderately thick discs (very thin, radiatively efficient discs are of
course much harder to simulate). Thinner discs being denser, the
Alfv\`en speed is reduced in their inner part, and maintaining the
causal connection between the magnetized gas in the plunging region and
the disc should be more difficult. 
Indeed, Afshordi \& Paczynski (2002), by studying a numerical model of a
steady state, {\it thin} and isothermal accretion disc, argued
that the NT inner boundary condition is appropriate, 
provided that the structure carrying the stress in the disc 
(likely the magnetic field) has a scaleheight much smaller that the
radius, i.e. is itself geometrically thin (see also Li, 2003). 

It is worth stressing here that the very existence (and the
variability properties) of 
broad and redshifted fluorescent Iron K$\alpha$ lines
in the spectra of accreting black holes, put constraints on the disc
models: the discs that produce them have to be cold (not too highly
ionized), optically thick to Compton scattering 
and illuminated by a strongly variable source of hard X-rays. Clearly,
non-radiative thick accretion flows, as those simulated by Hawley and
Krolik (2002), do not satisfy those conditions.

Therefore, beside the no-torque (NT) and the disc-torque (DT) boundary conditions
advocated, respectively, by Paczynski and collaborators and by Krolik
and collaborators, we envisage a third plausible
alternative, applicable to the case in which a
geometrically thin, cold disc is indeed present together with the
hotter coronal phase, as 
inferred from the detection 
of broad iron K$\alpha$ fluorescent lines. The proposed geometry for
the inner flow is the following:
a cold thin disc, with no torque at the inner boundary,
giving rise to the observed broad lines via reflection of hard X-ray photons,
is surrounded by a magnetically dominated \cite{mf01} corona. 
Magnetic field lines
that penetrate the ergosphere at high latitude above the equatorial
plane are connected to the coronal plasma, and the torque exerted by
the rotation of the inertial frames transports energy into the corona
(coronal thread\footnote{As our corona is not accreting,
  and no angular momentum conservation equation is discussed, it would be
  incorrect to talk of coronal torque.} case, CT). This is in practice
a generalization of the Blandford-Znajek \cite{bz77} mechanism, 
where the energy
extracted from the rotating hole instead of powering a relativistic
outflow is dissipated, via magnetic reconnection, in the corona.
The coronal emissivity profile can 
then be more centrally peaked than in the NT case, because of the 
excess field dissipation in its innermost part. 

\begin{figure*}
\psfig{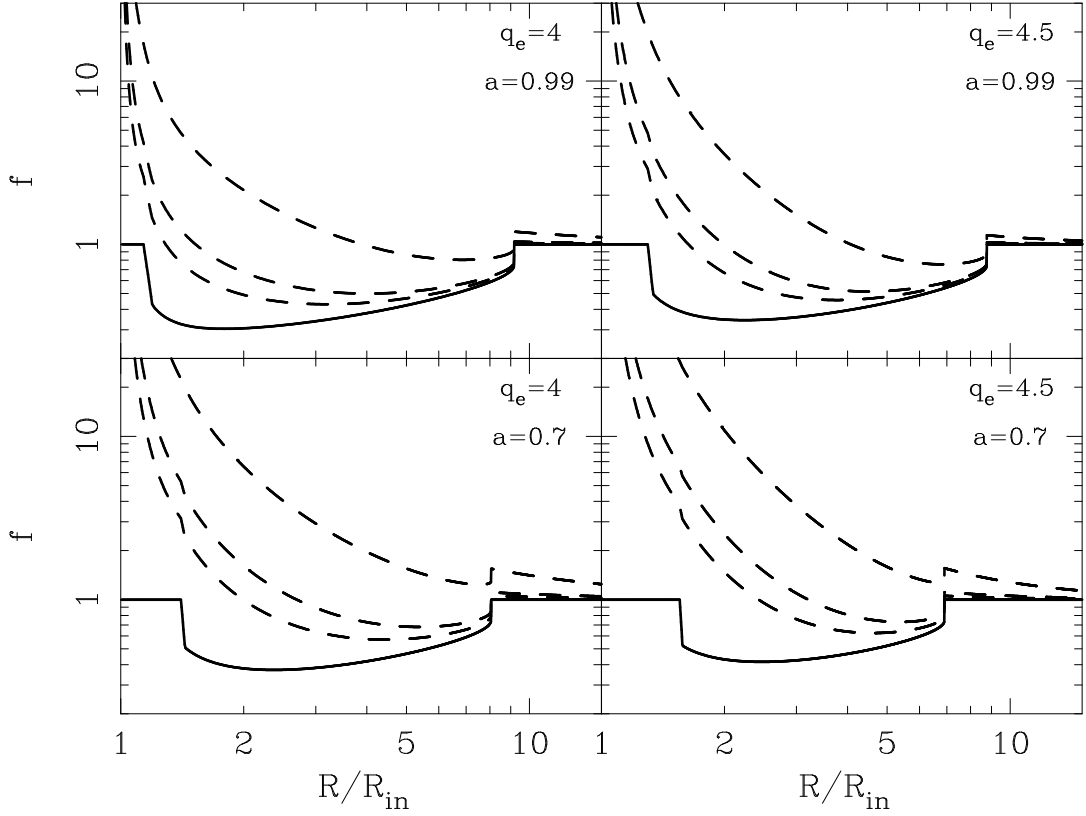}
\caption{The profiles of the coronal fraction $f(R)$ for $\Delta\epsilon=0$ (NT, thick solid lines) and 
for the case of extra energy dissipation in the corona (CT) 
with $\Delta\epsilon=0.05, 0.1, 0.5$ (dashed lines, $\Delta \epsilon$
increasing from bottom to top).
The dimensionless black hole spin $a=0.99, 0.7$ and the index of
coronal dissipation $q_e$ are indicated
in each different panel. 
The dimensionless accretion rate, defined as the ratio of the emerging
bolometric luminosity to the Eddington one has been kept fixed to 
$\dot m \equiv L_{rm bol}/L_{\rm Edd} = \epsilon(a) \dot M_0 c^2/ L_{\rm Edd}=
0.03$.}
\label{prof_ct}
\end{figure*}

In the following sections we treat separately the two extreme case of
a torque applied to the thin disc only (DT, section~\ref{sec_dt}) and
of a direct magnetic connection between the black hole and the coronal
plasma (CT, section~\ref{sec_ct}). First, we solve the equations for the
coupled accretion disc--corona system and derive the coronal
emissivity profiles in the two cases, then  (section~\ref{sec_lines})
we discuss their effects on the properties of the fluorescent
iron lines profiles.

\subsection{Torque exerted on the thin disc}
\label{sec_dt}
We follow here the argument presented by Krolik (1999) and
adopt the formalism of Agol \& Krolik (2000), in particular,
we assume nonzero stress at the marginally stable orbit, so that the
flux at the disk surface in the fluid frame is not zero there.
The relativistic equations for the disc structure discussed in
Agol \& Krolik (2000) are modified here by the inclusion of a term that takes
into account the fraction of power that goes into the corona and
 supplemented with the coupling provided by Eq.~(\ref{eq_fr}). 
The corona, in this case, is coupled only to the underlying disc and
not to the free-falling gas, nor to the black hole horizon. 

Using this boundary condition, we have, for the locally generated 
radiation flux emerging from the disc:
\begin{equation}
\label{eq_flux}
F^{\rm DT}(x)=\frac{3}{8\pi}\frac{GM\dot M_0 (1-f)}{R^3}\left(\frac{x_{\rm
    ms}^{3/2}{\cal C}_{\rm ms}^{1/2}\Delta\epsilon}{{\cal C}(x)x^{1/2}}+R_{\rm
    R}^{\rm NT}\right).
\end{equation}
Here $\Delta\epsilon$ is the additional radiative efficiency relative
to the one computed in terms of binding energy at $r_{\rm ms}$ (the
usual disc efficiency, $\epsilon(a)$), $x=R/R_g=Rc^2/GM$ and ${\cal C}(x)$ 
is one of the auxiliary functions that
incorporate relativistic corrections to disk equations (from Novikov
and Thorne 1973, see also the Appendix). $R_{\rm F}^{\rm NT}$ is a factor
that represent the relativistic correction factors in the NT case 
\cite{nt73}; it goes to zero at $R_{\rm ms}$.
The corresponding angular momentum conservation equation reads 
\begin{equation}
\label{eq_tor}
-\int T_{r\phi}^{DT}(z) dz = \frac{\dot M_0 \Omega_{\rm K}(x)}{2\pi}\left(\frac{x_{\rm
    ms}^{3/2}{\cal D}_{\rm ms}^{1/2}\Delta\epsilon}{{\cal C}(x)x^{1/2}}+R_{\rm
    T}^{\rm NT}\right)
\end{equation}
where in the above formula $R_{\rm T}^{\rm
  NT}$ represent the torque relativistic correction factor
  (Novikov and Thorne, 1973), analogous to the one included in the
  flux formula.

With the help of Eqs.(\ref{eq_flux}) and (\ref{eq_tor}), 
we can again extract
the disc structure equations and find the solutions for the coronal
fraction as a function of $R$, for different $\Delta\epsilon$. This is
shown in  Figure~\ref{prof_dt}, for different black hole dimensionless
spin parameters and
 for three different values of 
$\Delta\epsilon$, between 0.05 ($\la \epsilon(0)\simeq 0.06$) and
0.5 ($\ga \epsilon(1)\simeq 0.42 $). 
We have fixed the dimensionless accretion rate, defined as the ratio of 
the bolometric luminosity to the Eddington one 
$\dot m\equiv L_{\rm b,acc}/L_{\rm Edd} = \epsilon(a) \dot M_0 c^2/ L_{\rm Edd}=
0.03$.

As expected, when a torque is applied to the geometrically
thin, optically thick disc only, 
the excess dissipation at the inner boundary reduces the
ratio $P_{\rm gas}/P_{\rm rad}$ and, consequently, $f$.
The more energy is extracted from the plunging region and transferred
to the disc, the less coronal emission is
produced in the innermost region of the accretion flow.

As it is well known \cite{ak00,li02}, when extra torque is applied in
the innermost region of an accretion disc, the emissivity profile {\it
  of the disc itself} becomes much steeper. Then, if the coronal
fraction does not decrease too rapidly with decreasing distance from
the hole, 
the coronal emissivity may be peaked, too. However,
the {\it intensity} of the non-thermal coronal flux compared with the
disc one is much reduced in the DT case, up to the point where coronal
emission becomes negligible. In practice,
 given our prescription for the disc--corona coupling, a torque
applied at the inner boundary of the thin disc only, has the effect of
quenching the  coronal activity in the innermost part of the accretion flow.
In section~\ref{sec_lines}  we will show a few quantitative measures of the amount
of peakedness of the coronal emissivity profile, and discuss its
observable effects on the iron K$\alpha$ line profiles for different
black hole spin parameters and $\Delta \epsilon$. Before that, we
examine next the effect of extra energy deposition on the corona itself.

\begin{figure*}
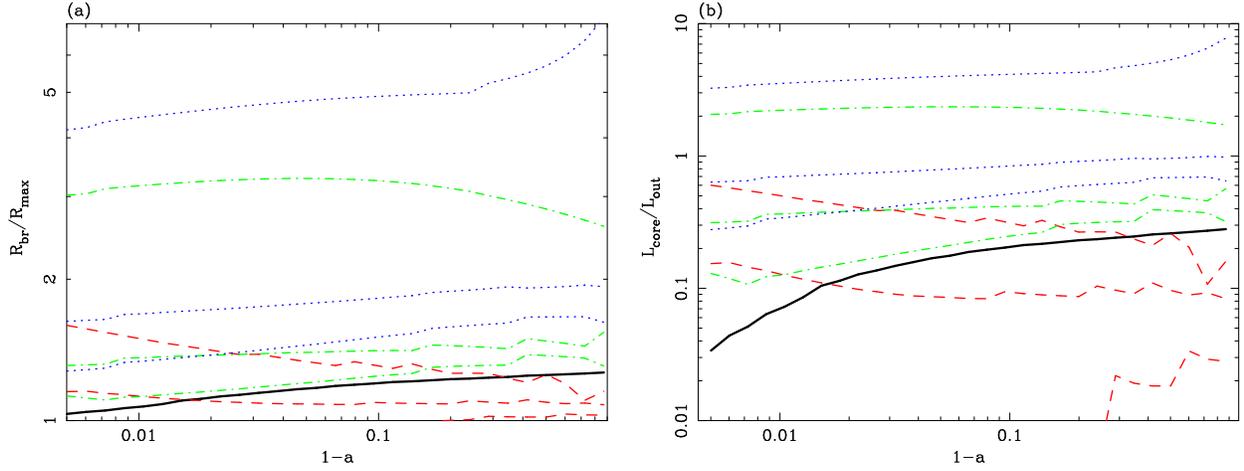

\begin{tabular}{cc}
\psfig{angle=270,width=0.45\textwidth,file=radii.ps}&
\psfig{angle=270,width=0.45\textwidth,file=line_ratio.ps}\\
\end{tabular}
\caption{Plots of the ratio $\xi$ of the core radius $R_{\rm br}$ to the radius of
  maximal coronal emissivity $R_{\rm max}$ (panel a), and of 
the ratio of the core coronal luminosity $L_{\rm core}$
  (computed from $R_{\rm max}$ to $R_{\rm br}$), to the outer one
  (from $R_{\rm br}$ to $R_{\rm out}=100 R_{\rm S}=200 GM/c^2$; panel
  b), both as functions of the black
  hole dimensionless angular momentum parameter $a$. 
The thick black solid lines represent values computed for $\Delta\epsilon=0$ (NT). 
Three disc torque cases are shown with (red) dashed lines for  
$\Delta\epsilon=0.05, 0.1$ and $0.5$, from bottom to top, respectively.
CT cases correspond to (green) dot-dashed lines ($q_e=4$) and 
  (blue) dotted lines ($q_e=4.5$), again going from 
$\Delta\epsilon=0.05, 0.1$ to $0.5$ from bottom to top. 
In all the calculations, we have fixed 
the dimensionless accretion rate to $\dot m=0.03$. 
}
\label{ratio}
\end{figure*}

\subsection{The black hole--corona connection}
\label{sec_ct}
The magnetically dominated plasma in the
low-density corona (where Alfv\`en speed may become relativistic) 
may indeed be the repository of most of the energy
extracted by electromagnetic processes from a spinning black hole 
\cite{kro99,bla01,sem02}. In this respect, we may consider the inner
part of the magnetically
dominated corona as the non-stationary analogous of the force-free 
magnetosphere envisaged by Blandford \& Znajek (1977). 
The extra energy may also come from an increased
coronal activity produced by the strongly magnetized plasma in the plunging
region \cite{gam99,kro99}, where efficient reconnection events may 
disconnect the free-falling plasma from the disc 
(thus reducing the effective torque at the inner disc boundary), while
simultaneously increasing the energy available for coronal heating.
 The time-averaged effect of such a process of energy extraction and 
deposition (whose physical modeling, involving the electrodynamics of
a reconnecting field in a curved background spacetime, 
is beyond the aim of this paper)
on the radial profile of the hard X-ray radiation illuminating the
underlying cold disc is the generation of a central cusp.

In order to consider the effect of a black hole--corona connection, with the
consequent increase of coronal emissivity in the simplest possible
way, we apply a similar argument to that presented in the previous
subsection.
The effect of the excess energy dissipation in the coronal region,
 is parametrized by the same factor 
$\Delta\epsilon$ representing the extra radiative efficiency of the
accreting gas with respect to the standard value $\epsilon(a)$. 
However, in this case, we leave unchanged the disc
equation for the locally generated flux:
\begin{equation}
F(x)=\frac{3}{8\pi}\frac{GM\dot M_0 (1-f^*(R))}{R^3}R_{\rm
    R}^{\rm NT},
\end{equation}
as well as the angular momentum conservation equation:
\begin{equation}
-\int T_{r\phi}(z) dz = \frac{\dot M_0 \Omega_{\rm K}(x)}{2\pi}R_{\rm
    T}^{\rm NT},
\end{equation}
which are therefore both identical to the no-torque case.

Then, we modify the expression for the local coronal flux
by simply adding a term with a power-law profile:
\begin{equation}
F_{\rm c}^{\rm CT}(R)=\frac{3 G M \dot M_0}{8
  \pi}\left(\frac{f^*(R)}{R^3}
\frac{{\cal Q}
  {\cal B}}{ {\cal C}^{1/2}}+ {\cal K}_e R^{-q_{\rm e}}\right)\; {\rm ergs}\; {\rm cm}^{-2} \; {\rm s}^{-1}.
\end{equation}

The additional term is normalized in such a way that the extra coronal
luminosity $L_e=2\int^{\infty}_{R_{\rm in}}2\pi R Q_e(R) dR$, where we
have defined $Q_e(r)=(3/8\pi)GM \dot M_0 K_e R^{-q_{\rm e}}$, is equal to 
$\Delta\epsilon  \dot M_0 c^2$, and thus we have 
\begin{equation}
{\cal K}_e=\frac{4}{3 R_{\rm S}} (q_{\rm e} -2) R_{\rm in}^{(q_{\rm e}-2)}.
\Delta\epsilon
\end{equation}

Here we need to introduce a new definition of the fraction of power
released into the corona. We define $f(R)=F^{\rm CT}_{\rm
  c}(R)/Q(R)=f^*(R)+f^{\rm e}(R)$, where now $f^*(R)$ 
is the (usual) fraction of
dissipated accretion energy deposited into the corona (bound to be
less than unity, see section~\ref{centr}), 
that enters the disc structure equations, while
$f^{\rm e}(R)$ is the additional coronal energy in magnetic field lines
connected to the plunging gas at moderate-to-high latitude.
As opposed to the DT case, here the more the energy extracted from the
black hole (i.e. the bigger $\Delta\epsilon$), the larger the coronal
fraction $f(R)$. Indeed, as expected, in the innermost disc region, the
coronal fraction, i.e. the ratio of the energy dissipated in the
optically thin, hot, magnetically dominated corona to the total
gravitational energy locally dissipated by the accretion flow, $Q$,
may be larger than unity. This is clearly shown in  
Figure~\ref{prof_ct}, where, again, we show the profiles of $f(R)$ for
both highly and moderately spinning black holes, $a=0.99$ and $a=0.7$,
for two different values of the index $q_{\rm e}$ and
for three different values of $\Delta\epsilon$, between 0.05 and 0.5.
Globally, we can define the integrated fractions of power released into
the corona in the following way:
\begin{equation}
<f^*>\equiv \frac{2\int^{\infty}_{R_{\rm in}}2\pi R f^*(R) Q(R) dR}
{2\int^{\infty}_{R_{\rm in}}2\pi R Q(R) dR}\equiv\frac{L_{\rm c}}{L_{\rm b,acc}}.
\end{equation}
Then we get, for the global value of the fraction of power released
into the corona, $<f>$:
\begin{equation}
<f>=\frac{L_e+L_{\rm c}}{L_{\rm
    b,acc}}=\frac{\Delta\epsilon}{\epsilon(a)}+<f^*>,
\end{equation}
from which we deduce that the extra term due to the energy transported from inside $R_{\rm ms}$
and released into the corona, $<f^{\rm e}>$ is given by the ratio 
$\Delta\epsilon/\epsilon(a)$ of the extra radiative efficiency to the
standard (accretion) one.

\section{Effects on the Fe K$\alpha$ line profiles}
\label{sec_lines}
In order to estimate the effect of the coronal emissivity profiles on
the shape of the fluorescent Iron K$\alpha$ line produced by the
underlying cold, optically thick disc, we have approximated the
computed $F_{\rm c}(r)$ with a double power-law (see Fabian et al.,
2002, for an application to the observation of MCG--6-30-15), assuming that the
inner radius corresponds to the radius of maximal coronal emissivity
($R_{\rm in}=R_{\rm max}$). 
For the inner core, extending from $R_{\rm max}$
out to a break radius $R_{\rm br}$, we have approximated the 
peaked profile with an emissivity index $\beta=4$, the same for all
the computed profiles. 
Thus, for the normalized coronal flux we have
\begin{eqnarray}
F_{\rm c}(R)&=&\left(\frac{R_{\rm max}}{R}\right)^{4}\; \; \; \; {\rm for}\;\;
R_{\rm max}<R<R_{\rm br} \nonumber \\
F_{\rm c}(R)&=& \left(\frac{R_{\rm
      max}}{R_{\rm br}}\right)^{4}\left(\frac{R_{\rm br}}{R}\right)^{\beta_2}\;\;\;\; {\rm for}
\;\; R_{\rm br}<R<R_{\rm out}.
\end{eqnarray} 
By fitting such a profile to the calculated $F_{\rm c}$, we find the
core radius $R_{\rm br}$ and the emissivity index, $\beta_2$, of the outer,
flatter part.  

Also, returning radiation \cite{cun76} 
has to be taken into account\footnote{In fact, also the vertical
structure of the illuminating corona may become important, and a full
general relativistic treatment of light bending effects \cite{mmk02}
for realistic coronal geometries should be
relevant to exactly determine the line shape in the most extreme
cases.}. This can be done
approximately, by using the analytic fitting formula for the fraction
of the returning radiation to the locally generated one of Agol \&
Krolik (2000). For $R_{\rm out} \ga 10 R_{\rm in}$ such a fraction, $R_{\rm
  ret}$ can be approximated with a constant, that we denote 
$R_{\rm ret,\infty}$. The returning flux then scales as $r^{-3}$:
\begin{equation}
F_{\rm ret}(R)=R_{\rm ret,\infty}Q(R_{\rm out}) 
\left(\frac{R_{\rm out}}{R}\right)^3. 
\end{equation}
We can then introduce  the integrated
  luminosity from the core region $L_{\rm core}=4\pi
  \int^{R_{\rm br}}_{R_{\rm max}} 
[F_{\rm c}(R)+F_{\rm ret}(R)] R dR$ (and,
  correspondingly,
that of the outer region $L_{\rm
  out}=4\pi \int^{R_{\rm out}}_{R_{\rm br}} [F_{\rm c}(R))+F_{\rm ret}(R)] R dR$). 
By integrating these expressions, we get, for
 the relative luminosity of the two regions, (for $\beta_2
\neq 2$)
\begin{equation}
\label{norm}
\frac{L_{\rm core}}{L_{\rm out}}=\frac{\beta_2-2}{2}
\left(\frac{(\xi^2-1)+2 A \xi^4 \chi^3 (\xi-1)}
{(1-\chi^{(2-\beta_2)})+(p-2)A \xi^4 \chi^2 (\chi-1)}\right), 
\end{equation}
where $\xi=R_{\rm br}/R_{\rm max}$, $\chi=R_{\rm out}/R_{\rm br}$
and $A=R_{\rm ret,\infty} Q(R_{\rm out})$. 

Both  $\xi$ and $L_{\rm core}/L_{\rm out}$ can be used as measures
of the peakedness of the coronal emissivity profile. The values of
these two quantities determines the profile of the fluorescent iron
line produced by the reflection of the illuminating coronal radiation
on the cold disc. 
In Figure~\ref{ratio}a we plot the ratio $\xi=R_{\rm br}/R_{\rm max}$ of
the break radius to the inner radius as a 
function of the black hole dimensionless spin $a$, for different
values of $\Delta \epsilon$, for the cases of DT, CT and NT.
In Figure~\ref{ratio}b we plot the ratio $L_{\rm core}/L_{\rm
  out}$ of the luminosity of the coronal core region to the luminosity of the
  outer coronal region. These figures show that the inner core in the DT case
is {\it less} prominent than in the NT case for $a<0.99$ and
$\Delta\epsilon<0.1$. Only for $\Delta\epsilon \ga 0.5$ do the coronal
emissivity profiles in the DT case become substantially peaked. On the
other hand, for the coronal-thread case (CT), already for
$\Delta\epsilon = 0.05$ is the ratio $L_{\rm core}/L_{\rm out}$ larger
than in the NT case. Its value increases rapidly for increasing
$\Delta\epsilon$ and $q_{\rm e}$.
 
\begin{figure*}
\psfig{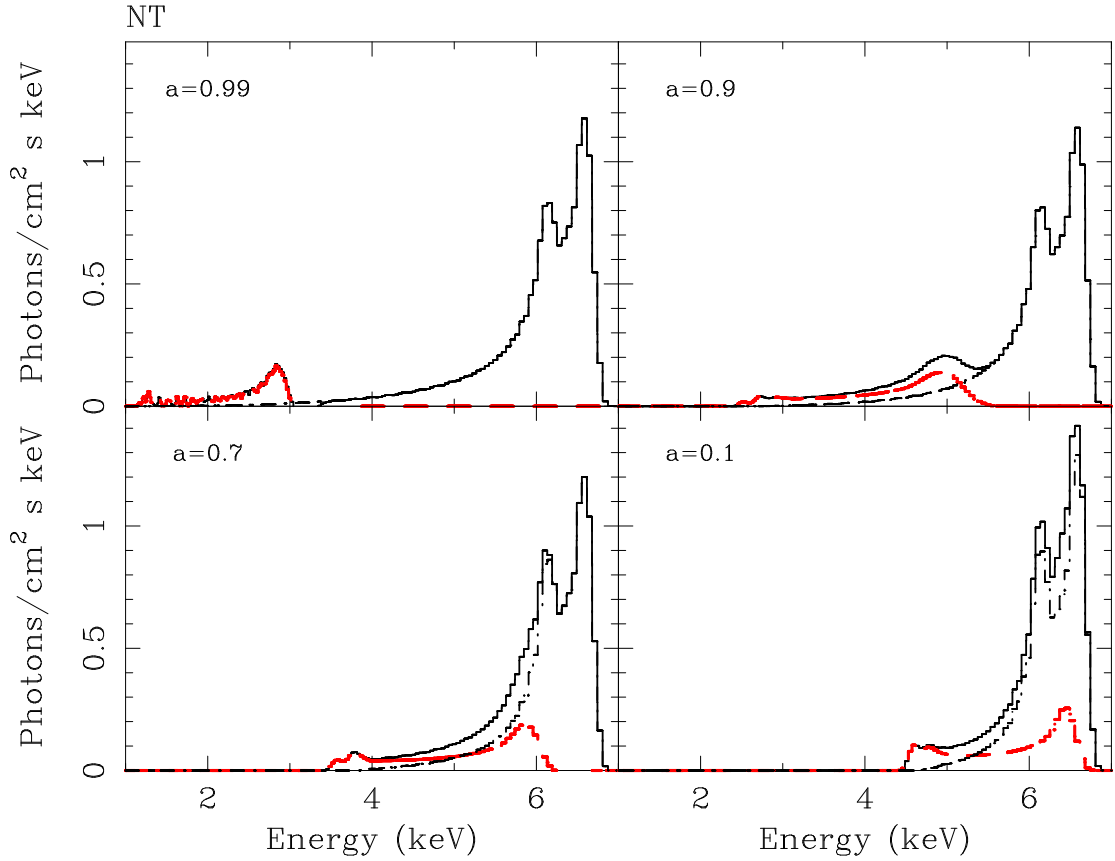}
\caption{The line profiles calculated from Laor model in the
  no-torque at the inner boundary case (NT, $\Delta\epsilon=0$), with two
  components: a core, between the inner radius and the core radius,
  with steep emissivity profile ($\beta=4$, red dashed lines), and an
  outer region, from the core radius to the outer radius, with a more
  shallow emissivity profile. The
  solid lines represent the total line emission. 
  Four different values of the black hole spin are shown.}
\label{lineallNT}
\end{figure*}

\begin{figure*}
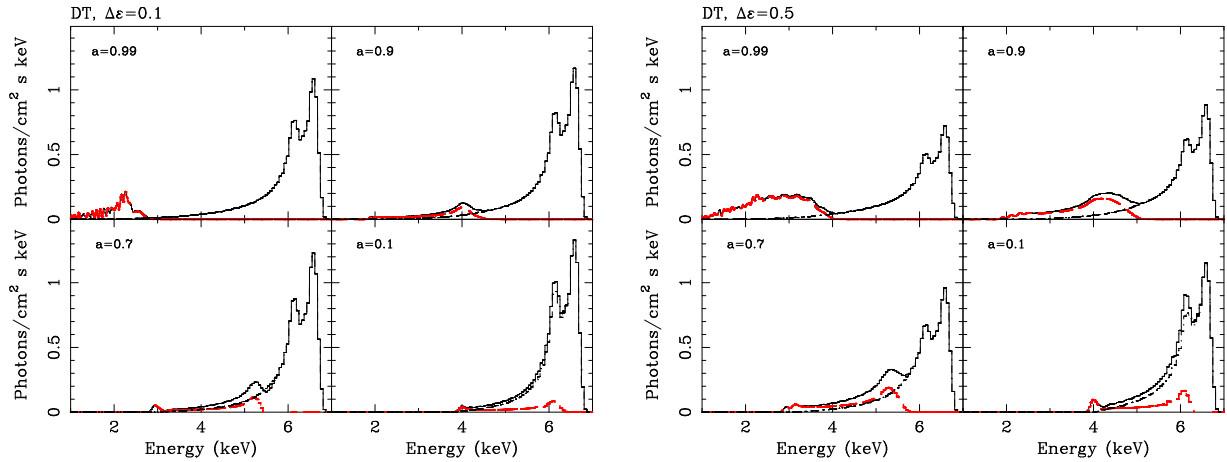

\begin{tabular}{cc}
\psfig{angle=270,width=0.45\textwidth,file=line_fig_dt_de01.ps}&
\psfig{angle=270,width=0.45\textwidth,file=line_fig_dt_de05.ps}\\
\end{tabular}
\caption{As in Figure~\ref{lineallNT}, in the 
disc-torque at the inner boundary case (DT), for 
$\Delta\epsilon=0.1$ (left) and $0.5$ (right) and four values of the
spin parameter $a$.}
\label{lineallDT}
\end{figure*}

\begin{figure*}
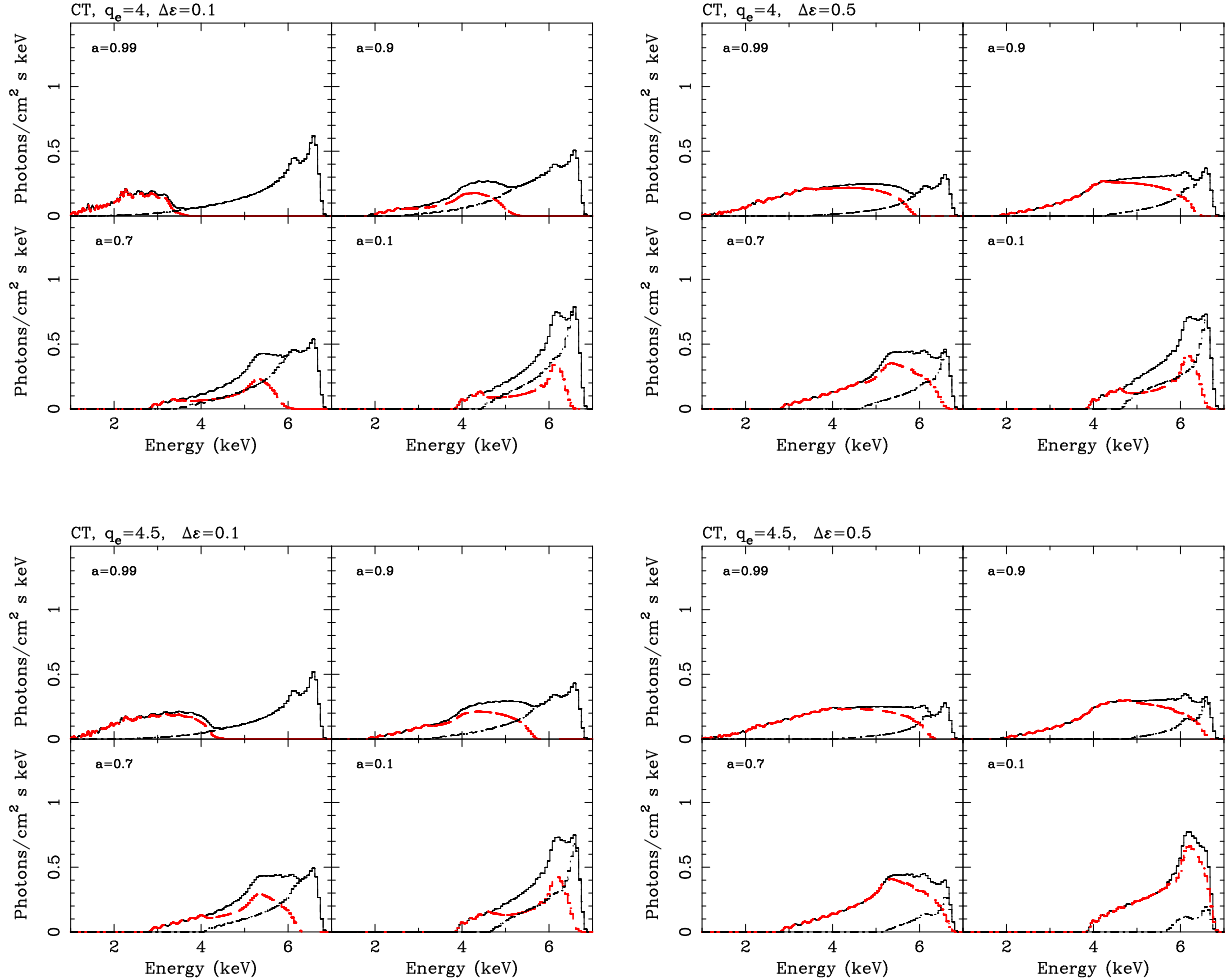

\begin{tabular}{cc}
\psfig{angle=270,width=0.45\textwidth,file=line_fig_ct_de01.ps}&
\psfig{angle=270,width=0.45\textwidth,file=line_fig_ct_de05.ps}\\
\psfig{angle=270,width=0.45\textwidth,file=line_fig_ct45_de01.ps}&
\psfig{angle=270,width=0.45\textwidth,file=line_fig_ct45_de05.ps}\\
\end{tabular}
\caption{As in Figure~\ref{lineallNT}, in the 
corona-torque at the inner boundary case (CT), for 
$\Delta\epsilon=0.1$ (left) and $0.5$ (right), $q_{\rm e}=4$ (top) and
$q_{\rm e}=4.5$ (bottom), and four values of the
spin parameter $a$.}
\label{lineallCT}
\end{figure*}

To compute the line profiles we have adopted the Laor (1991) model
as implemented in XSPEC. Each line is the sum of two components, one
produced by the core, the other by the outer region. 
The model for each component has the following parameters: inner
and outer radius, emissivity index $\beta$, line energy, inclination
angle and overall normalization. In all cases, the line energy was 
fixed to 6.4 keV, the emissivity index of the core region to
$\beta=4$, the outer radius to $100 R_{\rm S}$ 
and the inclination angle to 30$^\circ$. 
The variable parameters that determine the line shape are: 
the inner radius (coincident with $R_{\rm max}$),
the break radius, $R_{\rm br}$ (which is simultaneously 
the outer radius of the core
component and the inner radius of the outer component), the emissivity index of the outer
component, $\beta_2$, and the normalizations of the two components. 
These have been fixed by imposing $N_{\rm core}+N_{\rm out}=1$ and taking
$N_{\rm core}/N_{\rm out}=L_{\rm core}/L_{\rm out}$ as computed with
the help of Eq.~\ref{norm}. 
Table~\ref{tablelines} shows the value of the parameters of the double
power-law fit to the coronal emissivity profiles that have been used as
inputs into the Laor
model used to produce the line profiles shown in
figures~\ref{lineallNT}, \ref{lineallDT} and \ref{lineallCT}.

\begin{table}
  \caption{Physical parameters (inner radius, $r_{\rm max}$; core
    radius, $r_{\rm br}$, both in units of Schwarzschild radii;
    outer emissivity index, $\beta_2$; 
    and relative
    normalizations of the core and outer component), obtained from a double power-law fitting of the
    coronal emissivity profiles for different values  of the black hole
    dimensionless spin parameter $a$, of the extra radiative
    efficiency $\Delta\epsilon$, and of the inner boundary
    condition (T). In the fits, the core emissivity index has been
    fixed to $4$ and the outer radius to $100 R_{\rm S}$.}
\label{tablelines}
  \begin{tabular}{@{}ccccccccc@{}}
   \multicolumn{4}{c}{Model parameters} &  \multicolumn{5}{c}{Fit parameters}\\
  T & $a$ & $\Delta\epsilon$  & $q_{\rm e}$ & $r_{\rm max}$ & $r_{\rm
   br}$ & 
$\beta_2$ & $N_{\rm core}$ & $N_{\rm out}$ \\ [2pt]
\hline \\
NT & $0.99$ & $0$ & -- & $0.873$ & $0.897$ & $2.06$ & $0.068$ & $0.932$ \\
NT & $0.9$ & $0$ & -- & $1.49$ & $1.78$ & $1.97$ & $0.169$ & $0.831$ \\
NT & $0.7$ & $0$ & -- & $2.33$ & $2.88$ & $1.90$ & $0.194$ & $0.806$ \\
NT & $0.1$ & $0$ & -- & $4.27$ & $5.40$ & $1.81$ & $0.219$ & $0.781$ \\
\hline \\
DT & $0.99$ & $0.1$ & -- & $0.734$ & $0.826$ & $2.11$ & $0.112$ & $0.888$ \\
DT & $0.9$  & $0.1$ & -- & $1.15$ & $1.25$ & $2.06$ & $0.086$ & $0.914$ \\
DT & $0.7$  & $0.1$ & -- & $1.67$ & $1.83$ & $2.01$ & $0.088$ & $0.912$ \\
DT & $0.1$  & $0.1$ & -- & $2.79$ & $3.02$ & $2.02$ & $0.077$ & $0.923$ \\
DT & $0.99$ & $0.5$ & -- & $0.734$ & $1.09$ & $2.30$ & $0.331$ & $0.669$ \\
DT & $0.9$  & $0.5$ & -- & $1.15$ & $1.51$ & $2.23$ & $0.241$ & $0.759$ \\
DT & $0.7$  & $0.5$ & -- & $1.67$ & $2.11$ & $2.25$ & $0.210$ & $0.790$ \\
DT & $0.1$  & $0.5$ & -- & $2.79$ & $3.22$ & $2.29$ & $0.138$ & $0.862$ \\
\hline \\
CT & $0.99$ & $0.1$ & $4$ & $0.734$ & $0.999$ & $2.57$ & $0.267$ & $0.733$ \\
CT & $0.9$  & $0.1$ & $4$ & $1.15$ & $1.62$ & $2.85$ & $0.293$ & $0.707$ \\
CT & $0.7$  & $0.1$ & $4$ & $1.67$ & $2.40$ & $3.02$ & $0.306$ & $0.694$ \\
CT & $0.1$  & $0.1$ & $4$ & $2.79$ & $4.30$ & $3.11$ & $0.361$ & $0.639$ \\
CT & $0.99$ & $0.5$ & $4$ & $0.734$ & $2.34$ & $2.58$ & $0.689$ & $0.311$ \\
CT & $0.9$  & $0.5$ & $4$ & $1.15$ & $3.75$ & $2.74$ & $0.700$ & $0.300$ \\
CT & $0.7$  & $0.5$ & $4$ & $1.67$ & $5.06$ & $3.06$ & $0.679$ & $0.321$ \\
CT & $0.1$  & $0.5$ & $4$ & $2.79$ & $7.26$ & $3.44$ & $0.632$ & $0.368$ \\
CT & $0.99$ & $0.1$ & $4.5$ & $0.734$ & $1.24$ & $2.55$ & $0.411$ & $0.589$ \\
CT & $0.9$  & $0.1$ & $4.5$ & $1.15$ & $2.09$ & $2.88$ & $0.455$ & $0.545$ \\
CT & $0.7$  & $0.1$ & $4.5$ & $1.67$ & $3.20$ & $3.10$ & $0.484$ & $0.516$ \\
CT & $0.1$  & $0.1$ & $4.5$ & $2.79$ & $5.38$ & $3.31$ & $0.495$ & $0.505$ \\
CT & $0.99$ & $0.5$ & $4.5$ & $0.734$ & $3.28$ & $2.48$ & $0.780$ & $0.220$ \\
CT & $0.9$  & $0.5$ & $4.5$ & $1.15$ & $5.65$ & $2.58$ & $0.805$ & $0.195$ \\
CT & $0.7$  & $0.5$ & $4.5$ & $1.67$ & $8.79$ & $2.95$ & $0.823$ & $0.177$ \\
CT & $0.1$  & $0.5$ & $4.5$ & $2.79$ & $20.2$ & $3.49$ & $0.886$ & $0.114$ \\
\hline \\
\end{tabular}
%\end{minipage}
\end{table}

In Figure~\ref{lineallNT} we show the
composite lines we obtain for the NT case for four values of the black hole
spin ($a=0.1, 0.7,0.9,0.99$). The red dashed profiles show the line components
produced from the core region, while the black, dot-dashed ones those
from the outer regions. Obviously, the higher the
black hole spin, the more red-shifted is the core-component, which
appear as a small excess at low energies.
In Figure~\ref{lineallDT} we show the profiles derived in the CT case
for two values of the parameter $\Delta\epsilon$. Comparison with the
NT case demonstrate that, for $\Delta\epsilon=0.1$, the redshifted core 
components are substantially weaker and more redshifted (because
$r_{\rm max}$ is smaller in the DT case, see Table~\ref{tablelines}). 
For larger $\Delta\epsilon$,
the broad red shoulder produced by the core of the coronal emission
becomes more evident, especially for rapidly spinning black holes.
Finally, in Figure~\ref{lineallCT} we show the effect of the modified coronal
emissivity profile in the case of black hole--corona connection (CT),
for two different values of the power-law index of the radial profile
of the extra energy dissipation, $q_{\rm e}$. 
As we expect from the increase of $L_{\rm
  core}/L_{\rm out}$ with increasing $\Delta\epsilon$ shown in
Fig.~\ref{ratio}b, the strongly redshifted component of the line,
associated with the coronal core is more and more prominent.
 
The line profiles obtained in the CT with an extra efficiency of the
order of 0.1 (and in the DT case for much larger values of
$\Delta\epsilon$), 
are indeed consistent with those observed by {\it
  XMM Newton} in the X-ray spectrum of the Seyfert 1 galaxy
MCG--6-30-15 \cite{fab02}, indicating the possibility that dissipation
of rotational energy of the black hole tapped by electromagnetic
processes has indeed been observed. The calculation we have performed
clearly demonstrate that the line profiles are more sensitive to energy of the
gas in the plunging region (and/or in the ergosphere) being
magnetically transported and  
dissipated into the corona rather than into the disc, in the sense that the
same amount of extra luminosity (additional to the bolometric
accretion one) produces much stronger effects in the CT case than in the DT one.   

\section{Discussion}
\label{sec_discuss}
The calculations we have presented in order to include the effects of
different inner boundary conditions into observables are obviously
oversimplified, and we discuss below the main uncertainties that may
affect our conclusions. 

First of all, we only considered stationary
solutions of the accretion equations. As clearly demonstrated by a
number of simulations \cite{ra01,hk02} the nature of the accretion
flow at the inner boundary is intrinsically time dependent, and the
angular momentum transport there, which is mediated by the filaments
with the higher relative magnetic field \cite{ra01}, can vary strongly
on the dynamical timescale. However the line profiles we aim at
modeling are usually derived integrating the observed radiation over
many dynamical timescales. In fact, a direct comparison of
time-resolved spectra could shed some light on the complex time
variability of the inner accretion flow.

Also, we apply a (generalized) $\alpha$ stress model to describe
angular momentum transport, which may be questionable if, as we
assume, the MRI-driven MHD turbulence saturates due to buoyancy of
magnetic flux tubes \cite{bp99}. But our approach is essentially
phenomenological, and the work presented here aims at understanding the
most important pieces of information hidden in
marginal effects on observable physical quantities 
(for example, in the iron line profiles)
rather than solve a complex theoretical issue. 
In this respect, the advantage of the
simplifying $\alpha$ prescription is to allow a relatively
straightforward interpretation of the observations.

Another reason of uncertainties is the identification of the
inner edge of the reflecting cold disc (the ``reflection edge'', Krolik \&
Hawley 2002) with the radius of the marginally stable orbit around the
central black hole. As discussed in \cite{kh02}, the true reflection
edge, i.e. the place where the integrated surface density of the disc
has declined enough that the disc becomes effectively transparent, may
depend on the accretion rate (and on the fraction of power released
into the corona), if magnetic stresses in the inner disc are such that
the radial inflow accelerates smoothly inward rather that having
a sharp transition at $R_{\rm ms}$. Numerical simulations of cold,
thin, radiative discs at different accretion rates are clearly needed
in order to better verify these hypothesis.

Finally, contrary to the cases of NT and DT inner boundary conditions, where,
for a given small number of assumptions, a full self-consistent analytic
model can be solved for the disc--corona system, the coronal-thread
hypothesis is admittedly more speculative: we do not have a specific
model for how the energy can be extracted from the black hole and
deposited there (but see discussion at the beginning of
section~\ref{sec_ct}),
and we do not take into account the problem  of 
angular momentum transfer in the corona (our coronae are effectively
non accreting). Once again, given all the theoretical uncertainties, 
we believe it is more appropriate to adopt the simplest 
assumption about how the energy
extracted from the black hole energizes the corona 
(see section~\ref{sec_ct}) and explore its observational consequences.    

Given all the above uncertainties, it is obvious that all three kinds
of boundary condition we have adopted are highly idealized. None of
them is likely to fully describe the physical condition in the
innermost part of a turbulent disc surrounded by a magnetically
dominated corona. Nonetheless, their predictions for the intensity of
the coronal emission compared with the quasi-thermal disc one, for the
coronal emissivity profiles and for the shape of the fluorescent Iron K$\alpha$
lines are qualitatively very different in the three cases 
and relatively easy to distinguish.
That is what makes the use of such idealized models worthwhile.
In fact, the current observations provides us with enough information
to constrain the models. The shape of the iron line is very
sensitive to the coronal emissivity profiles. Both the spin of the
black hole (through the amount of gravitational redshift the line 
photons suffer) and the extra efficiency $\Delta \epsilon$ (through
the relative intensity of the red wing of the line, which fixes the
emissivity index and size of the core region) can in principle be
measured. Simultaneous multi-wavelength observations can set limits on
the relative fraction of coronal versus disc emission, therefore
helping discriminate between the CT and DT cases. The full picture can
then emerge after a careful comparison with existing models of all
these pieces of informations.

\section{Conclusions}
\label{sec_conc}
Our work demonstrates that detailed spectral analysis of the most
broadened and redshifted iron K$\alpha$ lines from accreting black
holes (and of their variability) is a fundamental tool with which to
constrain both the geometry of the innermost part of the disc-corona
system and the nature of the inner boundary condition of relativistic
accretion flows.

The emissivity profile to be considered in analysing the data, to which the
line shape is very sensitive, must be that of the illuminating flux,
and therefore of {\it coronal} radiation, rather than that of the disc.
When the generation of a magnetically dominated corona is included 
self-consistently in the relativistic equations for standard accretion
discs, the coronal emissivity profiles are usually strongly
peaked in the inner part, as far as a portion of the disc is radiation
pressure dominated. The extent of such an inner coronal ``core''
depends crucially on the inner boundary condition.
Here we have explored with our
calculation three different (idealized) prescriptions:
a classical no-torque at the inner boundary condition;
a magnetic connection to the plunging region of the  
{\it cold, geometrically thin disc} only
(disc-torque) and a magnetic connection to the plunging region of the
    {\it vertically extended, magnetically dominated corona} only
    (coronal-thread).
We show that, while in the NT case indeed  the coronal emission is 
more centrally peaked than the disc one, this effect is much enhanced
in the CT case, with the strongest deviation from the standard
emissivity profile appearing in the innermost coronal region for
larger black hole spin and larger amount of energy tapped from the
plunging region (and/or the black hole itself). On the other hand,
when the torque is applied directly to the thin disc (DT case),
although the emissivity profiles may become again more centrally
peaked (for quite large values of $\Delta\epsilon$),
the coronal power is much reduced, and we expect the emission to be
completely dominated by the quasi-thermal radiation coming from the
optically thick disc.

By approximating the exact coronal emissivity profiles with a broken
power-law, we have shown how the detailed line profiles are sensitive
to the location of the break radius and to the relative normalization
of the inner and outer illuminating regions. These parameters depend
strongly on the black hole spin, on the amount of extra energy
deposited into the corona, and on the nature of the
inner boundary condition itself.
The recently observed extremely broad and skewed fluorescent lines
in the X-ray spectra of the Seyfert 1 galaxy MCG--6-30-15 
\cite{wil01,fab02}, may therefore be read as 
an indication that the extraction of energy
from spinning black holes is most likely mediated by the magnetic corona
rather than by the geometrically thin, optically thick accretion disc.

\section*{Acknowledgments}
We thank Henk Spruit for helpful discussion. 
ACF thanks the Royal Society for support.

\appendix
\section{Relativistic disc structure equations with modified viscosity
  law}
Following Novikov \& Thorne (1973), we define
\begin{eqnarray}
{\cal A}&=& 1+a^2x^{-4}+2a^2 x^{-6}\nonumber \\
{\cal B}&=& 1+a x^{-3}\nonumber \\
{\cal C}&=& 1-3 x^{-2}+2 a x^{-3}\nonumber \\
{\cal D}&=& =1-2 x^{-2}+a^2 x^{-4}\nonumber \\
{\cal E}&=& =1+4 a^2 x^{-4}-4a^2 x^{-6}+3a^4 x^{-8}\nonumber \\
{\cal Q}&=& \frac{1+ax^{-3}}
          {x(1-3x^{-2}+2ax^{-3})^{1/2}} [x-x_0-1.5a \ln(x/x_0)
            \nonumber \\ &&-
\frac{3(x_1-a)^2}{x_1(x_1-x_2)(x_1-x_3)} \ln \frac{(x-x_1)}{(x_0-x_1)}-
\frac{3(x_2-a)^2}{x_2(x_2-x_1)(x_2-x_3)} \ln \frac{(x-x_2)}{(x_0-x_2)}-
\frac{3(x_3-a)^2}{x_3(x_3-x_1)(x_3-x_2)} \ln \frac{(x-x_3)}{(x_0-x_3)}
]
      \nonumber \\
\end{eqnarray}
where $x=\sqrt{R c^2/GM}$, $x_0=\sqrt{R_{\rm ms}c^2/GM}$, and
$x_1=2\cos(1/3 \cos^{-1}(a) -\pi/3)$, $x_2=2\cos(1/3
\cos^{-1}(a)+\pi/3)$,
$x_3=-2\cos(1/3 \cos^{-1}(a))$.

\begin{itemize}
\item[(a)]{When magnetic stresses are proportional to the
geometric mean of gas
and total pressure \cite{m03},
in the case of radiation pressure dominated discs, we obtain the
following expressions for disc scaleheight, pressure, density  
 (both calculated at the disc mid-plane), temperature and ratio of the radiation to gas pressure, respectively:
\begin{eqnarray}
\label{eq_struc_rad_cor_mod}
h &\simeq& 10.6 \dot m (1-f) 
{\cal A}^{2}{\cal B}^{-3}{\cal C}^{1/2}{\cal D}^{-1}{\cal E}^{-1}{\cal Q}
\\
P & \simeq & 3.15 \times 10^{18} (\alpha m)^{-8/9} r^{-8/3} 
\dot m_0^{8/9} 
{\cal A}^{-16/9}{\cal B}^{8/9}{\cal C}^{-4/9}{\cal E}^{8/9}{\cal Q}^{8/9}
\; {\rm dyne}\\
\rho  & \simeq & 1.29 \times 10^{-4} (\alpha m)^{-8/9} r^{3/9} 
\dot m_0^{-10/9} (1-f)^{-2} 
{\cal A}^{-26/9}{\cal B}^{4}{\cal D}^{5/9}{\cal E}^{13/9}{\cal Q}^{-10/9}
\; {\rm g \; cm}^{-3} \\
T & \simeq & 1.86 \times 10^8  (\alpha m)^{-2/9} r^{-2/3} 
\dot m_0^{2/9}  
{\cal A}^{-2/9}{\cal D}^{-1/9}{\cal E}^{1/9}{\cal Q}^{2/9}
\; {\rm K} \\
\frac{P_{\rm rad}}{P_{\rm gas}} &\simeq& 7.97 \times 10^5 (\alpha m)^{2/9} r^{-7/3} 
\dot m_0^{16/9} (1-f)^{2} 
{\cal A}^{20/9}{\cal B}^{-4}{\cal D}^{-8/9}{\cal E}^{-10/9}{\cal Q}^{16/9},
\end{eqnarray}
where the dimensionless parameters are defined as: $r=R/R_{\rm S}$,
$\dot m_0=0.06 \dot M_0 c^2/L_{\rm Edd}$ and $m=M/M_{\odot}$.}

\item[(b)]{
In the gas pressure dominated part of the disc, instead we have:
\begin{eqnarray}
\label{eq_struc_gas_cor}
h & \simeq & 2.02 \times 10^{-2} \xi^{-3/32} (\alpha m)^{-1/10}
r^{21/20} \dot m_0^{1/5} (1-f)^{1/10}  
{\cal A}{\cal B}^{-6/5}{\cal C}^{1/2}{\cal D}^{-3/5}{\cal E}^{-1/2}{\cal Q}^{1/5}
\\
P & \simeq & 1.88\times 10^{18} \xi^{3/32} (\alpha m)^{-9/10}
r^{-51/20} \dot m_0^{4/5}(1-f)^{-1/10}  
{\cal A}^{-1}{\cal B}^{9/5}{\cal D}^{2/5}{\cal E}^{1/2}{\cal Q}^{-4/5}
\; {\rm dyne}\\
\rho  & \simeq & 20.3  \xi^{11/32} (\alpha m)^{-7/10} r^{-33/20} (1-f)^{-3/10} 
\dot m_0^{2/5}
{\cal A}^{-1}{\cal B}^{3/5}{\cal D}^{-1/5}{\cal E}^{1/2}{\cal Q}^{2/5}
\; {\rm g \; cm}^{-3} \\
T & \simeq & 5.6 \times 10^8  \xi^{-1/4} (\alpha m)^{-1/5} r^{-9/10} 
\dot m_0^{2/5} (1-f)^{1/5} 
{\cal B}^{-2/5}{\cal D}^{-1/5}{\cal Q}^{2/5}
\; {\rm K} \\
\frac{P_{\rm rad}}{P_{\rm gas}} & \simeq & 1.31 \times 10^2
\xi^{-35/32} (\alpha  m)^{1/10} r^{-21/20} \dot m_0^{4/5} (1-f)^{9/10}
{\cal A}{\cal B}^{-9/5}{\cal D}^{-2/5}{\cal E}^{-1/2}{\cal Q}^{4/5}. 
\end{eqnarray}
The value $\xi\simeq 1/3$, the numerical that takes into account the 
difference 
in the vertical density profile that should enter in the definition of
the disc scaleheight for a gas pressure dominated disc, 
can be found by imposing continuity of
all physical quantities at the boundary between the two regions (see
Merloni, 2001).
}
\end{itemize}
\bsp

\label{lastpage}


\begin{thebibliography}{}

\bibitem[Abramowicz et al. 1988]{abr88}
Abramowicz, M. A. Czerny, B., Lasota, J.-P. \& Szuszkiewicz, E., 
1988, ApJ, 332, 646.

\bibitem[Abramowicz \& Kato 1989]{ak89}
Abramowicz, M. A. \& Kato, S., 1989, ApJ, 336, 304.

\bibitem[Afshordi \& Paczynski 2002]{ap02}
Afshordi, N., Paczynski, B., 2002, ApJ, submitted, astro-ph/0202409

\bibitem[Agol \& Krolik 2000]{ak00}
Agol, E. \& Krolik, J. H., 2000, ApJ, 528, 161.

\bibitem[Balbus \& Hawley 1991]{bh91}
Balbus, S. A. \& Hawley, J. F., 1991, ApJ, 376, 214.

\bibitem[Balbus \& Hawley 1998]{bh98}
Balbus, S. A. \& Hawley, J. F., 1998, Rev. Mod. Phys. 70, 1.

\bibitem[Balbus \& Papaloizou 1999]{bp99}
Balbus, S. A. \& Papaloizou, J. C. B., 1999, ApJ, 521, 650.

%\bibitem[\protect\citename{Ballantyne, Ross \& Fabian }2001]{brf01}
%Ballantyne D. R., Ross R. R. \& Fabian A. C., 2001, MNRAS, 327, 10.

\bibitem[Blaes \& Socrates 2001]{bs01}
Blaes, O. \& Socrates, A., 2001, ApJ, 553, 987.

\bibitem[Blandford 2001]{bla01}
Blandford, R. D., 2001, Prog. Theor. Phys. Supp., 143, 182.

\bibitem[Blandford \& Znajek 1977]{bz77}
Blandford, R. D. \& Znajek, R. L., 1977, MNRAS, 179, 433.

\bibitem[Cunningham 1976]{cun76}
Cunningham, C. T., 1976, ApJ, 208, 534.

\bibitem[Burm 1985]{bur85}
Burm, H., 1985, A\&A, 143, 389.

\bibitem[Fabian et al. 2002]{fab02}
Fabian, A. C. et al., 2002, MNRAS, 335, L1.

\bibitem[Fabian et al. 2000]{fab00}
Fabian, A. C., Iwasawa, K., Reynolds, C. R. \& Young, A. J., 2000,
PASP, 112, 1145.

\bibitem[Galeev, Rosner \& Vaiana 1979]{grv79}
Galeev, A. A., Rosner, R. \& Vaiana, G. S., 1979, ApJ, 229, 318.

\bibitem[Gammie 1999]{gam99}
Gammie, C. F., 1999, ApJL, 522, L57.

\bibitem[Haardt \& Maraschi 1991]{hm91}
Haardt, F. \& Maraschi, L., 1991, ApJL, 380, L51.

%\bibitem[Haardt, Maraschi \& Ghisellini 1994]{hmg94}
%Haardt, F., Maraschi, L. \& Ghisellini, G., 1994, ApJL, 432, L95.

\bibitem[Hawley \& Krolik 2002]{hk02}
Hawley, J. F. \& Krolik, J. H., 2002, ApJ, 566, 164

\bibitem[Krolik 1999]{kro99}
Krolik, J. H., 1999, ApJL, 515, L73  

\bibitem[Krolik \& Hawley 2002]{kh02}
Krolik, J. H. \& Hawley, J. F., 2002, ApJ, 573, 754

\bibitem[Laor 1991]{lao91}
Laor, A., 1991, ApJ, 376, 90.

\bibitem[Li 2002]{li02}
Li, X.-L., 2002, ApJ, 567, 463.

\bibitem[Li 2003]{li03}
Li, X.-L., 2003, astro-ph/0212456


%\bibitem[Livio, Ogilvie \& Pringle 1999]{lop99}
%Livio, M., Ogilvie, G. I. \& Pringle, J. E., 1999, ApJ, 512, 100.

%\bibitem[Machida, Hyashi \& Matsumoto 2000]{mach00}
%Machida, M., Hayashi, M. R. \& Matsumoto, R., 2000, ApJ, 532, L67.

\bibitem[Martocchia, Matt \& Karas 2002]{mmk02}
Martocchia, A., Matt, G. \& Karas, V., 2002, A\&A, 383, L23

\bibitem[Merloni 2001]{mer01}
Merloni, A., 2001, PhD thesis.

\bibitem[Merloni 2003]{m03}
Merloni, A., 2003, MNRAS, submitted.

\bibitem[Merloni \& Fabian 2001]{mf01}
Merloni, A. \& Fabian, A. C., 2001, MNRAS, 321, 549.

\bibitem[Merloni \& Fabian 2002]{mf02}
Merloni, A. \& Fabian, A. C., 2002, MNRAS, 332, 165.

\bibitem[Miller et al. 2002]{mil02}
Miller, J. M. et al., 2002, ApJL, 570, L69.

\bibitem[Muchotrzeb \& Paczy\'nski 1982]{mp82}
Muchotrzeb, M. \& Paczy\'nski, B., 1982, AcA, 32, 1.


\bibitem[Novikov \& Thorne 1973]{nt73}
Novikov, I. D. \& Thorne, K. S., 1973, in Black Holes, ed. C. De Witt
\& B. De Witt (New York: Gordon \& Breach), 343.

\bibitem[Page \& Thorne 1974]{pt74}
Page, D. \& Thorne, K. S., 1974, ApJ, 191, 499.

\bibitem[Reynolds \& Armitage 2001]{ra01}
Reynolds, C. S. \& Armitage, P. J., 2001, ApJL, 561, L81.

\bibitem[Sakimoto \& Coroniti 1981]{sc81}
Sakimoto, P. J. \& Coroniti, F. V., 1981, ApJ, 247, 19.

\bibitem[Sakimoto \& Coroniti 1989]{sc89}
Sakimoto, P. J. \& Coroniti, F. V., 1989, ApJ, 342, 49.

\bibitem[Semenov et al. 2002]{sem02}
Semenov, V. S., Dyadechkin, S. A., Ivanov, I. B.,  Biernat, H. K., 
2002, Physica Scripta, 65, 13.

\bibitem[\protect\citename{Shakura \& Sunyaev }1973]{ss73}
Shakura, N. I. \& Sunyaev, R.A., 1973, A\&A, 24, 337.


\bibitem[Stella \& Rosner 1984]{sr84}
Stella, L. \& Rosner, R.,  1984, ApJ, 277, 312.

\bibitem[\protect\citename{Svensson \& Zdziarski }1994]{sz94}
Svensson, R. \& Zdziarski, A. A., 1994, ApJ, 436, 599.

%\bibitem[\protect\citename{Szuszkiewicz }1990]{szu90}
%Szuszkiewicz, E., 1990, MNRAS, 244, 377

\bibitem[\protect\citename{Taam \& Lin }1984]{tl84}
Taam, R. E. \& Lin, D. N. C., 1984, ApJ, 287, 761

\bibitem[Turner, Stone \& Sano 2002]{tss02}
Turner, N. J., Stone, J. M. \& Sano, T.,  2002, ApJ, 566, 148.

\bibitem[Wilms et al. 2001]{wil01}
Wilms, J., Reynolds, C. S., Begelman, M. C., Reeves, J., Molendi, S.,
Staubert, R. \& Kendziorra, E., 2001, MNRAS, 328, L27.

\end{thebibliography}
\end{document}